\documentclass{article}
\usepackage{arxiv}
\usepackage[a4paper,width=17.4cm]{geometry}
\usepackage{amssymb}
\usepackage{amsfonts}
\usepackage{graphicx} % Required for inserting images
\usepackage{amsmath}
\usepackage[utf8]{inputenc}
\usepackage{hyperref}
\usepackage{capt-of}
\usepackage{ragged2e}
\usepackage{layouts}
\usepackage{cite}
\usepackage{xspace}
\usepackage[many]{tcolorbox}
\tcbset{
    sharpish corners, % better drop shadow
    boxrule = 0pt,
    toprule = 4.5pt, % top rule weight
    enhanced,
    parbox=false,
    capture=minipage,
    breakable,
    fuzzy shadow = {0pt}{-2pt}{-0.5pt}{0.5pt}{black!35} % {xshift}{yshift}{offset}{step}{options} 
}
\usepackage{float}
\usepackage{siunitx}

\urldef{\urlSinus}\url{6_Sinus_rhythm/frag/614_C_N_933_8s_frag}
\urldef{\urlVF}\url{1_Dangerous_VFL_VF/frag/426_C_VF_454_5s_frag}

\newtcolorbox[auto counter]{bbb}[2][]{
enhanced, fonttitle=\bfseries,
title={Box \thetcbcounter: #2},
#1
}
\usepackage{afterpage}
\newcommand*\mycircled[1]{\raisebox{.5pt}{\textcircled{\raisebox{-0.3pt}{\scalebox{0.75}{#1}}}}}

\usepackage{setspace}
\newcommand{\methods}{Methods\xspace}

\title{Synchronized Hearts, Desynchronized Brains: A Nonlinear Dynamics Perspective for \mbox{Function and Failure}}

\usepackage[ranges,superaftercomma,nobreak]{autoaffil}

\newcommand{\MPIDS}{\autoaffil{Max Planck Institute for Dynamics and Self-Organization, G\"ottingen, Germany}}
\newcommand{\IDCS}{\autoaffil{Institute for the Dynamics of Complex Systems, University of G\"ottingen, G\"ottingen, Germany}}
\newcommand{\MBExC}{\autoaffil{Cluster of Excellence “Multiscale Bioimaging: from Molecular Machines to Networks of Excitable Cells” (MBExC), University of Göttingen, Göttingen, Germany}}
\newcommand{\MtL}{\autoaffil{Max Planck School Matter to Life, Heidelberg, Germany}}
\newcommand{\CIDBN}{\autoaffil{G\"ottingen Campus Institute for Dynamics of Biological Networks,  G\"ottingen, Germany}}
\newcommand{\UMGNeuropath}{\autoaffil{Institute of Neuropathology, University Medical Center Göttingen}}
\newcommand{\UGOEXRay}{\autoaffil{Institute for X-Ray Physics, University of G\"ottingen, G\"ottingen, Germany}}
\newcommand{\UMGPharmTox}{\autoaffil{Institute of Pharmacology and Toxicology, University Medical Center G\"ottingen, Germany; University of G\"ottingen, G\"ottingen, Germany}}
\newcommand{\DZHKGOE}{\autoaffil{German Center for Cardiovascular Research (DZHK), partner site Lower Saxony, G\"ottingen, Germany}}
\newcommand{\DZNEGOE}{\autoaffil{German Center for Neurodegenerative Diseases (DZNE), G\"ottingen, Germany}}
\newcommand{\DZKJGOE}{\autoaffil{German Center for Child and Adolescent Health (DZKJ), G\"ottingen, Germany}}
\newcommand{\ITMP}{\autoaffil{Fraunhofer Institute for Translational Medicine and Pharmacology (ITMP), G\"ottingen, Germany}}
\newcommand{\CIDAS}{\autoaffil{Campus Institute Data Science (CIDAS), G\"ottingen, Germany}}
\newcommand{\UMGCardPhys}{\autoaffil{Institute of Cardiovascular Physiology, University Medical Center Göttingen}}
\newcommand{\UMGCardPneu}{\autoaffil{Department of Cardiology and Pneumology, Heart Research Center Göttingen, University Medical Center Göttingen}}
\newcommand{\UMGPedAdoMed}{\autoaffil{Clinic for pediatric and adolescent medicine, University Medical Center Goettingen}}
\newcommand{\EKFZOptoTher}{\autoaffil{Else Kröner Fresenius Center for Optogenetic Therapies (EKFZ L2T), Göttingen, Germany}}
\newcommand{\UMGAudNeuro}{\autoaffil{Institute for Auditory Neuroscience and InnerEarLab, University Medical Center Göttingen, Göttingen, Germany}}
\newcommand{\UMGNeuroSensPhys}{\autoaffil{Department for Neuro- and Sensory Physiology, University Medical Center Göttingen, Göttingen, Germany}}
\newcommand{\IMPRSNeuro}{\autoaffil{IMPRS Program Neurosciences, Göttingen, Germany}}

\autoremark{*}{I.P., S.B. contributed equally}
\autoremark{\dagger}{P.B., J.Z., W.-H. Z, V.P. contributed equally}
\autoremark{\ddagger}{Corresponding Authors: \texttt{johannes.zierenberg@ds.mpg.de}, \texttt{viola.priesemann@ds.mpg.de}}

\autoauthor[*]{Irene Pellini}
\MPIDS
\IDCS
\MBExC
\MtL

\autoauthor[*]{Simon Bauer}
\MPIDS
\IDCS

\autoauthor{Seba Contreras}
\MPIDS
\IDCS

\autoauthor{Ruben Fernández Busnadiego}
\MBExC
\UMGNeuropath

\autoauthor{Anushka Deb}
\UMGNeuroSensPhys
\IMPRSNeuro

\autoauthor{Sophia Hartmann}
\UMGNeuroSensPhys

\autoauthor{Sayedeh Hussaini}
\MPIDS
\IDCS

\autoauthor{Dörthe M. Katschinski}
\UMGCardPhys

\autoauthor{Sarah Köster}
\MBExC
\MtL
\UGOEXRay
\DZHKGOE

\autoauthor{Stephan E. Lehnart}
\MBExC
\DZHKGOE
\UMGCardPneu

\autoauthor{Tobias Moser}
\MBExC
\UMGAudNeuro

\autoauthor{Andreas Neef}
\MPIDS
\IDCS
\CIDBN

\autoauthor{Ulrich Parlitz}
\MPIDS
\IDCS

\autoauthor{Silvio Rizzoli}
\MBExC
\UMGNeuroSensPhys

\autoauthor{Maria-Patapia Zafeiriou}
\MBExC
\DZHKGOE
\UMGPharmTox
\UMGPedAdoMed
\EKFZOptoTher

\autoauthor[\dagger]{Wolfram-Hubertus Zimmermann}
\MBExC
\MtL
\DZHKGOE
\UMGPharmTox
\DZNEGOE
\DZKJGOE
\ITMP
\CIDAS

\autoauthor[\dagger, \ddagger]{Johannes Zierenberg}
\MPIDS
\IDCS
\MBExC

\autoauthor[\dagger]{Philip Bittihn}
\MPIDS
\IDCS

\autoauthor[\dagger,\ddagger]{Viola Priesemann}
\MPIDS
\IDCS
\MBExC
\MtL
\CIDAS

\date{}

\renewcommand{\headeright}{ }
\renewcommand{\undertitle}{ }

\begin{document}
\maketitle

%%%%%%%%%%%% Glossary %%%%%%%%%%%%%%
% emergent dynamics (not response, not activity)
% electrically excitable cells

\begin{abstract}
Both the heart and the brain are built from electrically excitable cells, yet the emergent dynamics of these organs could not differ more. 
In the myocardium, a coherent traveling wave of action potentials ensures coordinated contraction; in the cerebral cortex, distributed, irregular spiking underlies efficient information processing and the generation of behavior. 
Here, we illuminate the principles that underlie the emergence of such distinct non-linear dynamics using concepts from complex systems.
We trace these principles across scales from ion channels and cellular action-potential generation, through cell-to-cell coupling mechanisms and network architecture, to organ-level dynamics and clinical phenotypes.
Designed to support either synchronized waves or desynchronized information processing, the heart and the brain can thus be understood as operating on opposite sides of a synchronization phase transition. 
Crossing this transition implies pathological dynamics in both organs:
Loss of synchrony in the heart produces arrhythmia; excess synchrony in the brain leads to epileptic seizures.
We discuss how adaptation mechanisms, including plasticity, modulation, remodeling, and learning, maintain the appropriate level of (de)synchronization. Overall, we provide a unified excitable‑system perspective, together with an educational notebook, to foster cross‑disciplinary insights into cardiac and neural function and pathology. 
\end{abstract}

% \maketitle

% \section*{Significance Statement}
% The heart and brain both build their function from networks of electrically excitable cells. However, their functions differ fundamentally: healthy hearts rely on synchronized waves of activity, while healthy brains process information through desynchronized, distributed activity. Here, we compare the two organs side by side to identify key ingredients that shape emergent dynamics and, in turn, function. From a nonlinear dynamics perspective, we find the heart and brain on opposite sides of a synchronization transition. Crossing this transition induces pathological dynamics: loss of synchrony can cause cardiac arrhythmia; gaining excess synchrony can cause epileptic seizures. From our unifying perspective, we identify open questions that could guide future work at the intersection of cardiology, neuroscience, and nonlinear dynamics.

\section{Introduction}
%(Johannes)
\label{sec:intro}

Already in the 18th century, Luigi Galvani recognized the importance of electrical activity in physiology~\cite{galvani_viribus_1791}. 
His work paved the way for the later identification of electrically excitable cells as the building blocks underlying both cardiac contraction and neural information processing. 
In excitable cells, ion fluxes across the membrane maintain a negative resting potential; when this membrane potential is perturbed by a supra-threshold excitation, an electrical signal---a so-called action potential---is generated and propagates to excite or inhibit neighboring cells either electrically (heart) or biochemically (brain).
By generating such cascades, impulses and information can propagate through tissue in a coordinated manner.

Mathematically, cardiac and neuronal cells can be represented as excitable units that give rise to diverse emergent collective phenomena, such as coherent traveling waves that characterize cardiac tissue or distributed, ongoing activity that characterizes the brain. 
Generating the correct emergent activity is key to healthy organ function.
However, healthy function can be compromised by changes at the cellular or even subcellular level that can result in macroscopic organ-level pathologies, such as disruption of a coherent traveling wave causing  arrhythmia~\cite{jiang_current_2023,yazawa_using_2011, dick_arrhythmogenesis_2016} or excessive excitation causing epilepsy~\cite{chen_antisense_2024}. 
Such pathologies can be identified at the organ-level by non‑invasive recordings of collective activity: electrocardiograms (ECG) for the heart~\cite{einthoven_galvanometrische_1903} and electroencephalograms (EEG) for the brain~\cite{caton_electrical_1875,berger_uber_1929}.
Conversely, the microscopic defects underlying these conditions can be detected only with sophisticated but invasive microelectrodes~\cite{bretag_glass_2017} or optical imaging approaches~\cite{xu_voltage_2017}.
% This limits our understanding of pathologies and the development of medical treatments.
To overcome this experimental limitation and bridge microscopic mechanisms and macroscopic dysfunction, theoretical and computational approaches grounded in nonlinear dynamics are required.
These approaches can bridge the scales, from  micro-level modeling of membranes and cells, based on the classic membrane theories of Bernstein~\cite{bernstein_untersuchungen_1902} and Hodgkin–Huxley~\cite{hodgkin_quantitative_1952} (see Fig~\ref{fig:hodgkin-huxley}) to macro-level modeling of network topology and defects. 
Thereby, a thorough theoretical understanding can shed light on putative cellular origins of organ‑scale healthy and pathological collective dynamics.

%topic heart (function and disorder)
From a nonlinear dynamics viewpoint, cardiac function can be described as the coherent propagation of electromechanical waves through sheets of coupled cardiomyocytes with regular single-unit dynamics (sinus rhythm).
In a healthy heart, this wave is initiated regularly at the sinoatrial node, distributed by a specialized fiber network, and travels coherently across the cardiac tissue (myocardium) to produce the collective activity that underlies the mature ECG pattern and the coordinated contraction of the heart.
In this context, we define cardiac activity as synchronized in the sense that the excitation wavefront propagates coherently along the physiologically intended conduction path, maintaining spatial integrity and phase alignment, and thus ensuring coordinated contraction.
This definition quantifies deviations from healthy propagation: In cardiac arrhythmias, a broken, deformed, or reentrant wavefront deviates from this path and thus constitutes desynchronized, pathological activity.
This can happen when the traveling wavefront is altered, either intermittently or persistently, by structural and electrophysiological defects.
These defects range from cellular alterations to tissue‑level heterogeneities and can distort propagation, giving rise to a rich and challenging repertoire of spatiotemporal patterns~\cite{krogh-madsen_nonlinear_2012, debakker_ventricular_1993, pandit_rotors_2013,guevara_phase_1981,raad_arrhythmia_2021} (see Fig.~\ref{fig:spatiotemp}). 
Understanding how transitions from the regular sinus rhythm to these dysfunctional states emerge, informs the design of therapeutic strategies, including (i) low‑energy fibrillation control~\cite{ambrosi_termination_2011, luther_lowenergy_2011, suth_chaos_2024, hussaini_efficient_2024}, (ii) cardiac‑resynchronization therapy~\cite{vijayaraman_hispurkinje_2023,russo_acc_2025}, (iii) sub‑threshold cardiac‑contractility modulation~\cite{tint_oneyear_2023,rao_cardiac_2021}, (iv) antifibrotic interventions to improve heart function~\cite{ravassa_cardiac_2023, liu_cardiac_2025}, or (v) cardiomyocyte supplementation~\cite{zhang_trials_2025, jebran_engineered_2025}. 
These approaches aim at restoring the coordinated transmission of electrical signals necessary for efficient blood pumping based on the interpretation of pathological cardiac activity as a loss of synchronization in an excitable medium.

% topic: brain (function and disorder)
In contrast, brain function relies on a complex network of excitatory and inhibitory neurons that processes inputs in parallel streams, employing a myriad of mesoscopic dynamical states.
Such a variety of states originates from distributed activity with irregular single-unit dynamics, which is believed to support efficient encoding and processing of complex sensory inputs~\cite{stringer_spontaneous_2019, softky_highly_1993, mikulasch_where_2023}.
The resulting encoding is efficient in an information-theoretic sense because low correlations between units result in low redundancy and thereby higher information content of each pattern~\cite{mackay_information_2005, wibral_bits_2015}.
We thus define neural activity as asynchronous in the sense that neural firing shows low redundancy, which increases the accessible coding space to store and process many different patterns~\cite{mackay_information_2005, wilting_25_2019, brunel_dynamics_2000}.
Despite single-neuron statistics being predominantly asynchronous, coordination across brain areas emerges in the form of correlated fluctuations~\cite{fox_spontaneous_2007, stein_neuronal_2005, wilting_25_2019, aru_untangling_2015} and transient wave-like patterns across spatial and temporal scales~\cite{muller_cortical_2018, davis_spontaneous_2020, xu_interacting_2023}, which may reflect structured external inputs~\cite{stringer_highdimensional_2019, rudelt_signatures_2024} or specific physiological states such as REM sleep or attentive wakefulness~\cite{priesemann_neuronal_2013, sanchez-vives_multiscale_2025}.
As a consequence, there is a tradeoff between transient correlations for reliable communication and minimizing correlations for efficient representation~\cite{rudelt_signatures_2024}.
One hypothesis for how the brain achieves this tradeoff is that neural networks operate in the vicinity of a phase transition~\cite{tognoli_metastable_2014, priesemann_spike_2014, munoz_colloquium_2018, wilting_operating_2018a}. 
Such phase transitions can give rise to emergent pairwise correlations and can also explain the waxing‑and‑waning of oscillations~\cite{linkenkaer-hansen_longrange_2001, murray_hierarchy_2014, hasson_hierarchical_2015, palmigiano_flexible_2017, fontenele_criticality_2019, rudelt_signatures_2024}.
However, being positioned close to a phase transition carries the risk of crossing that transition, which implies excessive synchrony and potentially produces pathological, epileptic states~\cite{scharfman_neurobiology_2007, lehnertz_synchronization_2009, hagemann_assessing_2021, assenza_thalamocortical_2020, weng_macroscale_2020}. 
Interestingly, the role of synchronization is thus exactly opposite for the brain and the heart: the brain loses its function when synchrony becomes excessive, whereas the heart fails when the synchronization breaks down.

In this perspective, we provide an educational and cross-disciplinary perspective on both, heart and brain, because studying them side by side allows us to show how --- on each scale --- the design of these two organs is optimized for opposite collective dynamics according to their functional goals: traveling waves (synchronized, spatially coherent) for coordinated contraction versus distributed activity (asynchronous, low redundancy) for high information processing capacity.
Specifically, we review from a non-linear dynamics perspective the physiology, anatomy, network architecture, and cell‑cell coupling that differentiate the two organs (Sec.~\ref{sec:physiology}), then describe modeling approaches to the nonlinear dynamics of their common cellular building blocks (Sec.~\ref{sec:excitable}), and interpret their functional states as opposite sides of a synchronization phase transition (Sec.~\ref{sec:collective}) where crossing this transition produces pathological states (Sec.~\ref{sec:malfunction}). 
We complement this interpretation with an \textit{educational notebook} which allows one to explore the functional and pathological dynamics of the two organs hands-on~\cite{pellini_educational_2026}. 
Finally, we discuss how plasticity and modulation preserve function (Sec.~\ref{sec:plasticity}), and offer potential new research directions (Sec.~\ref{sec:openquestions}). 

\section{Basic physiology of heart and brain}
\label{sec:physiology}
% In the following, we examine the different scales at which interactions and signal transduction take place in the heart and brain (see Fig.~\ref{fig:fig1}). 
% These scales range from the organ level to the cellular network architecture and input structure, to cell-to-cell coupling, and inter-cell action potential generation. 
% Differences in all these scales support the respective organs' function and collective dynamics.
From a physiological viewpoint, the heart and brain have evolved an organ and network structure as well as dedicated cell-to-cell coupling mechanisms that support their specific functions. 
Despite their differences, however, the processes leading to electrical activity generation at the single-cell level are strikingly similar between the two organs. 
In the following, we examine the different scales at which interactions and signal transduction take place in the heart and brain (Fig.~\ref{fig:fig1}). 

\subsection{Organ level anatomy}
% heart

The heart is anatomically organized to ensure reliable blood pumping through adaptive regularity (Fig.~\ref{fig:fig1}B~\cite{zipes_cardiac_2017}). 
Regular contraction emerges from a hierarchical conduction system in which a primary pacemaking site, the sinoatrial node, initiates rhythmic electrical activity that propagates through four chambers (two atria and two ventricles) whose activation is temporally controlled by a specialized conduction system (the atrioventricular node, bundle of His, and Purkinje fibers~\cite{christoffels_development_2010}).
Functionally, this conduction hierarchy imposes a periodic organization on cardiac dynamics through the cardiac cycle. 
This cycle is initiated by the rhythmic activation of the sinoatrial node and is divided into diastole (relaxation) and systole (contraction).
During diastole, the chambers relax, and blood passively flows from the atria to the ventricles. 
Then, during atrial systole, the electrical impulse from the sinoatrial node spreads through the atria, causing atrial contraction that pushes further blood into the ventricles.
After a delay in the atrio‑ventricular node, the signal travels through the bundle of His and the Purkinje fibers to the ventricular myocardium. 
There, during ventricular systole, the electrical impulse triggers ventricular contraction, propelling deoxygenated blood into the pulmonary circulation (from the right ventricle to the lungs) to be oxygenated and oxygenated blood into the systemic circulation (from the left ventricle) to supply the body with it. 
Following ventricular systole, the chambers relax again, and the cycle repeats.  
Thus, the heart relies on an internally generated, rhythmic electrical input that is reliably propagated through its compartments to produce regular mechanical contraction.

The mammalian brain, on the other hand, is designed to process information in a distributed and functionally specialized manner.
It is organized into three major macro‑structures: the cerebral cortex, subcortical areas, and the cerebellum. 
The cortex, the outer layer of the brain, supports higher‑order functions such as perception, reasoning, language, and voluntary movement, and is conventionally divided into frontal, parietal, occipital, and temporal lobes~\cite{mtui_fitzgeralds_2020} (Fig.~\ref{fig:fig1}B). 
Beneath the cortex, subcortical structures --- including diencephalic regions such as the thalamus and hypothalamus, and limbic areas such as the amygdala and hippocampus --- regulate sensory relay, autonomic and endocrine control, emotion, and memory.
The cerebellum is essential for motor coordination, balance, and the fine‑tuning of movements. 
This macroscopic organization is hierarchical: within each region, there are specialized areas (for example, visual cortex V1 for edge detection, MT for motion, and the hippocampus CA1 for spatial navigation).
This specialization enables the brain to process multiple sensory inputs in parallel and to integrate them through recurrent interactions, thereby generating the spatially and temporally distributed, complex patterns of neural activity that underlie cognition and behavior.

\subsection{Network level architecture}
In the heart, coherent traveling waves are supported by the organization of the myocardium into sheets of cardiomyocytes that make up the thin myocardial walls of the atria and the thick walls of the ventricles~\cite{sachse_computational_2004, zipes_cardiac_2017} (Fig.~\ref{fig:fig1}C).
Within these quasi-two-dimensional sheets, cardiomyocytes are the main actors (an estimated 30-50\% of cells~\cite{lother_heterocellular_2023}) supported by non-myocytes such as endothelial cells, fibroblasts, smooth-muscle cells, leukocytes, and macrophages. 
How these different cell types interact is a field of active research~\cite{ceccato_defining_2020, wagner_cellular_2020, nicolas-avila_network_2020, hall_complex_2021, nicin_fibroblastmediated_2022}, not least because cardiac non-myocytes may alter conductivity~\cite{hulsmans_macrophages_2017, kohl_electrical_2005, quinn_electrotonic_2016} and form suitable targets for clinical diagnosis and therapy~\cite{gourdie_novel_2016, aghajanian_targeting_2019, zhao_targeting_2020, martins-marques_intercellular_2021, segers_autocrine_2021}. 
The working cardiomyocytes are tightly packed and mechanically anchored by resident fibroblasts and a collagen-rich extracellular matrix (ECM) that is secreted and continuously remodeled~\cite{bonnans_remodelling_2014}.
This ECM provides a robust yet flexible scaffold that supports both the efficient transmission of electric signals and of contractile forces by mediating the mechanical and electrical coupling between different cardiac sheets~\cite{carver_collagen_1991, creemers_molecular_2011}.
This inter-sheet coupling, alongside the local connections between cardiomyocytes called gap junctions, enables a single impulse generated by the sino‑atrial node to travel rapidly and coherently across the whole organ, producing the coordinated contraction that underlies the cardiac cycle. 
The reliability of wave propagation is further enhanced by the anisotropy of the connections, which are stronger along the longitudinal axis of the cardiomyocytes (the primary direction of wave propagation).

\afterpage{
\newsavebox{\footnotemarkheart}
\sbox{\footnotemarkheart}{\footnotemark}
\footnotetext{\url{https://commons.wikimedia.org/wiki/File:Shapes_of_the_cardiac_action_potential_in_the_heart.svg}}
\newsavebox{\footnotemarkbrain}
\sbox{\footnotemarkbrain}{\footnotemark}
\footnotetext{\url{https://commons.wikimedia.org/wiki/File:Gray728.svg}}
\begin{figure}[ht!]
    \centering
    \includegraphics{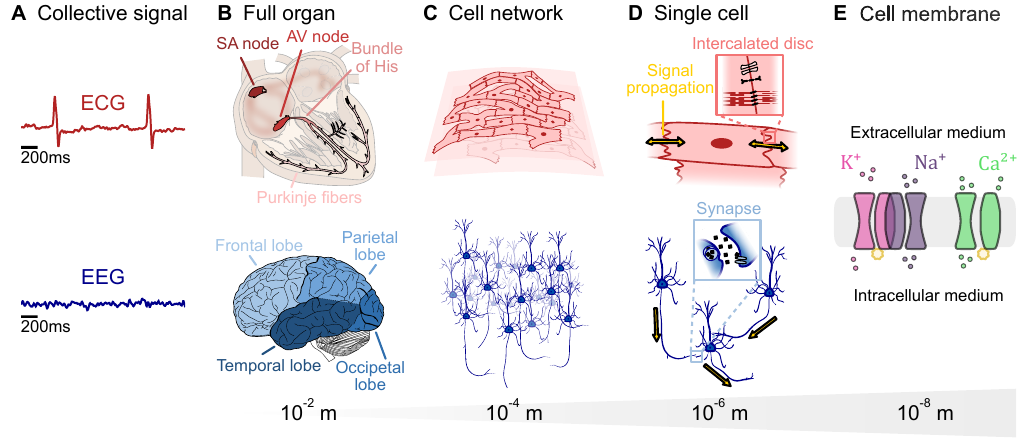}
    \caption{%
    \textbf{Both heart and brain are organs with specialized functions that depend on the collective dynamics of electrically excitable cells.}
    Their function, characterized by different collective signals (A), relies on the organization of the full organ (B), the specific architecture of the underlying cell networks (C), and the signal transmission between single cells (D). 
    In the heart, regular beating is ensured by traveling waves along a spatial structure with short-range interactions via bidirectional gap junctions and mechanotransduction. These interactions make the transmission immediate and reliable. 
    In the brain, information processing and input encoding occur in a complex network with a modular and hierarchical organization. 
    Transmission is directed, delayed, and weighted via chemical synapses that can change over time to enable learning.
    The underlying cells, cardiomyocytes in the heart and neurons in the brain, share a common electrical excitability through their cell membrane dynamics (E), which is regulated by the opening/closing of ion channels and pumps, which enable the formation of action potentials (see Fig.~\ref{fig:fig2}).\\
    Figure Sources: A created with data from Refs~\cite{manilo_ecg_2021, nemirko_ecg_2022, goldberger_physiobank_2000, andrzejak_indications_2001}; 
    B uses CardioNetworks~\usebox{\footnotemarkheart} (CCby3.0) and Wikimedia~\usebox{\footnotemarkbrain} (public domain).
   }
    \label{fig:fig1}
\end{figure}
}

In the brain, hierarchical information processing emerges from a complex network of neurons.
To form this network, neurons grow long “cables’’ (axons) that can form directed connections (synapses) to other neurons' cellular extensions (dendrites) (Fig.~\ref{fig:fig1}C). 
Whereas most axonal projections are short‑range to minimize wiring cost~\cite{cuntz_scaling_2012}, a substantial fraction of the axons also form long‑range connections that link distant cortical and subcortical regions, giving rise to a hierarchical, modular architecture~\cite{felleman_distributed_1991, meunier_modular_2010, zhang_geometric_2024}. 
Like in the heart, the neural tissue is embedded in an ECM that not only provides structural support~\cite{tewari_astrocytes_2024, bosiacki_perineuronal_2019} but also participates in long‑term memory consolidation~\cite{dityatev_extracellular_2003, frantz_extracellular_2010, tsien_very_2013}. 
Additionally, non‑neuronal glial cells (estimated as 60-90\% of brain cells~\cite{herculano-houzel_glia_2014}) perform essential trophic, signaling, electrically‑isolating, and immune functions~\cite{allen_glia_2009}. 
The resulting complex network is continuously remodeled by structural and synaptic plasticity, allowing learning and adaptation (see Sec.~\ref{sec:plasticity}). 
In contrast to the heart, whose local coupling sustains coherent wave propagation, the mixture of short- and long-range connections between neurons leads to asynchronous, multi-scale dynamics underlying efficient information processing in the brain.

\subsection{Cell-to-cell coupling}
% heart
Within cardiac sheets, cardiomyocytes are strongly and symmetrically coupled via multiple mechanisms to propagate activity from cell to cell.
%, supporting the reliable propagation of coherent activity wavefronts. 
This coupling is mediated by the intercalated disc, a junctional complex that connects two adjacent cells (Fig.~\ref{fig:fig1}D)~\cite{vermij_refining_2017, nielsen_intercalated_2023,zipes_cardiac_2017, zhao_intercalated_2019}. 
Within this complex, cells are coupled through different electrical (propagating the action potential), metabolic (e.g., ATP diffusion), and mechanical (propagating and buffering the contractile forces) pathways~\cite{vermij_refining_2017, nielsen_intercalated_2023}. 
The electric coupling is mainly mediated via direct intercellular ion exchange through gap junctions, namely low-resistance ion channels that form from two connexin hemichannels of adjacent cells~\cite{nielsen_intercalated_2023}, and through a high density of voltage-gated sodium channels (Nav) located near the junctions~\cite{gade_nav15_2025}.
In addition to this direct ion exchange, there is an electrical-field-mediated ephaptic coupling which indirectly induces action potentials in adjacent cardiomyocytes and can compensate, if needed, for impaired gap junctions~\cite{nielsen_intercalated_2023, cohen_immunocytochemical_1996}.
The mechanical coupling is mediated via desmosomes and adherens junctions~\cite{nielsen_intercalated_2023}. 
Desmosomes anchor cells together to support structure under strong contractile stresses, and adherens junctions couple the myofilaments of two cells, enabling transmission of contractile forces. 
These mechanical links appear earlier in development than electrical junctions~\cite{vreeker_assembly_2014}, suggesting that mechanical coupling alone can sustain coordinated impulse propagation in the embryonic heart.
It is interesting to note that all of these couplings are, in principle, bidirectional, but underlie the spatially directed propagation of excitation waves through the myocardium, where the anatomical structure and rhythmic activation in the healthy adult heart ultimately impose the desired activation sequence.

% brain
The coupling between neuronal cells, especially in the cortex, is largely delayed, directed to support directional information flow, and constantly adapted, resulting in asynchronous signaling. 
This comes from the most abundant type of coupling between neurons: chemical synapses (Fig.~\ref{fig:fig1}D).
In these synapses, the electrical activation of the sending (presynaptic) neuron triggers a delayed release of vesicles filled with specific signaling molecules (neurotransmitters) into the space between the presynaptic axon terminals and postsynaptic dendrites (synaptic cleft). 
There, the neurotransmitters bind to the postsynaptic receptors, modulating or mediating ion fluxes in the postsynaptic neurons, resulting in excitation or inhibition of the latter. 
For strong synapses, this signal can trigger an action potential in the postsynaptic neuron. 
For weak synapses, only the integration of many such signals from different synapses triggers the action potential. 
The strength of the synapse might be constantly adapted based on the electrical activity of the post-synaptic neuron as well as other neuromodulatory factors, a process called plasticity that increases the variability between synapses (Sec.~\ref{sec:plasticity}). 
Synaptic coupling is further shaped by glial cells: Astrocytes tightly regulate the extracellular ion and neurotransmitter environment, buffer K$^+$, clear glutamate, and release gliotransmitters, and we are only starting to understand their complementary role in adaptation and plasticity~\cite{perea_tripartite_2009, pannasch_astroglial_2011, allen_cell_2017}.
Besides chemical synapses, many brain regions exhibit bidirectional electrical coupling between neurons or between neurons and glia, akin to cardiac gap junctions~\cite{pereda_electrical_2014}. 
By providing low-resistance, almost instantaneous coupling between neighboring cells, these electrical synapses enable rapid local synchronization for, e.g., oscillations and rhythm generation~\cite{bennett_electrical_2004}.
This supports the notion that the delayed and directed coupling of chemical synapses facilitates asynchronous signaling, whereas the instantaneous and bidirectional coupling by gap junctions supports synchronization.

\subsection{Cellular action potentials}
Despite the drastic differences in organ structure and function, the generation of electrical activity in neurons and cardiomyocytes follows similar voltage traces (action potentials) and occurs, in first approximation, by the same basic ion movements across their membranes (Fig.~\ref{fig:fig1}E, Fig.~\ref{fig:hodgkin-huxley}). 
At rest, a negative membrane potential of -70 to -90 mV arises from the characteristic ionic composition of the cytoplasm - high intracellular \(\mathrm{K^{+}}\), and low intracellular \(\mathrm{Na^{+}}\) and \(\mathrm{Ca^{2+}}\). 
These concentrations are maintained by ion pumps, e.g., the \(\mathrm{Na^{+}}/\mathrm{K^{+}}\)-ATPase, and by ion channels that allow selective ionic fluxes.  
Electrical or chemical input from coupled cells can perturb this resting state and depolarize the membrane, increasing the membrane potential.
When depolarization crosses a threshold (around -50 mV), voltage‑gated \(\mathrm{Na^{+}}\) channels open, allowing \(\mathrm{Na^{+}}\) ions to rapidly flow inward along their electrochemical gradient. 
This flow produces the initial fast upstroke of the action potential. 
Subsequently, \(\mathrm{Na^{+}}\)‑channel inactivation, together with the slower activation of voltage‑gated \(\mathrm{K^{+}}\) channels, repolarizes the membrane, returning the membrane potential to its resting negative value. 
These inactivation and recovery dynamics generate a refractory period, during which additional stimuli cannot elicit another action potential, and the \(\mathrm{Na^{+}}/\mathrm{K^{+}}\) pumps gradually restore the ionic gradients. 
The successive phases of depolarization and repolarization characterize the action potential in both the heart and the brain.
This behavior arises from the intrinsic ability of the cell to remain quiescent under small perturbations yet respond explosively once a threshold is crossed. 
Such highly nonlinear, threshold-dependent dynamics are shared by cardiomyocytes and neurons and are referred to as excitability. 

\begin{figure}
    \centering
    \includegraphics{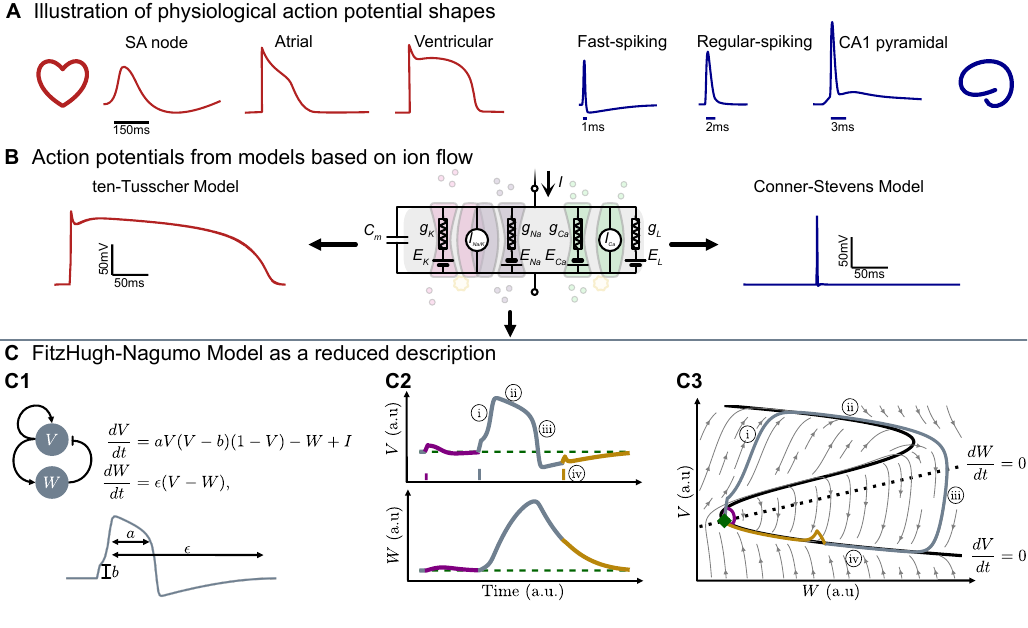}
    \caption{
    \textbf{Action potentials can be modeled as a nonlinear cell membrane response to external input.}
    \textbf{A.} Typical action potential shapes in the heart (modified from Ref.~\cite{ecgpedia_shapes_2011}) and brain (inspired by Ref.~\cite{bean_action_2007}) are functionally similar but differ in shape and duration (150 ms versus 1 ms, respectively). Longer action potentials facilitate synchronization.
    \textbf{B.} Realistic action potential shapes can be approximated by models that mimic ion-channel dynamics (Fig.~\ref{fig:hodgkin-huxley}~\cite{hodgkin_quantitative_1952,fatoyinbo_influence_2022}), leading to computational models for ventricular cardiomyocytes~\cite{tentusscher_model_2004} or cortical neurons~\cite{dayan_theoretical_2001}.
    \textbf{C.} An illustrative, reduced model is the 2-equation FitzHugh-Nagumo model that describes action potential generation with a non-linear self-activation of the membrane potential~$V$ and a slow recovery variable~$W$.
    The shape of the modeled action potential is determined by $b$, that sets the excitation threshold, $a$, that tunes the length of the depolarization, and ~$\epsilon$ that controls the total timescale of the action potential.
    Example time traces of $V$ and $W$ (C2) demonstrate the model's response to different pulse-like stimuli (small vertical bars), injected via the input current $I$. From left to right: A subthreshold stimulus failing to generate an action potential (purple) and decaying back to the resting state (green dashed line), a suprathreshold stimulus (gray), and a stimulus of the same strength but within the refractory period failing to generate an excitation (yellow). The suprathreshold stimulus initiates an action potential, which can be understood from the phase portrait (C3), with \mycircled{i} rapid initial depolarization, \mycircled{ii} plateau, \mycircled{iii} repolarization, and \mycircled{iv} hyperpolarization. Green diamond corresponds to the fixed point indicated as green dashed lines in panel C2.
    Parameters: $a=3$, $b=0.1$, $\epsilon=0.05$
    %The right-most panel shows a so-called phase portrait that visualizes the time traces in the $W$-$V$ plane together with the nullclines (solid $\dot{V}=0$ and dotted $\dot{W}=0$) and the local gradient field (arrows).
    }
    \label{fig:fig2}
\end{figure}

Based on these common ionic mechanisms, specific features of the action potential are tuned to meet organ-specific functional demands.  
In human atrial and ventricular cardiomyocytes, action potentials last between 200 and \SI{400}{\milli\second} (Fig.~\ref{fig:fig2}A left), far longer than the nearly instantaneous electrical coupling across gap junctions. 
This promotes successive activation along the myocardium, ensuring coherent signal propagation and preventing premature re-excitation. 
Importantly, the sustained duration stems from a calcium-generated, depolarized plateau phase that triggers cardiac contraction (excitation-contraction coupling). This is supported by a balance of long-lasting inward $\mathrm{Ca^{2+}}$ currents, late $\mathrm{Na^{+}}$ currents, and a minor membrane current carried by the $\mathrm{Na^{+}}$/$\mathrm{Ca^{2+}}$ exchanger.
The $\mathrm{Ca^{2+}}$ influx triggers calcium-induced calcium release from the sarcoplasmic reticulum. This internal signal amplification produces a robust rise in intracellular $\mathrm{Ca^{2+}}$ that activates cardiac contraction~\cite{zipes_cardiac_2017}.
Compared to atrial and ventricular cardiomyocytes, sinoatrial node pacemaker cells do not rely on the excitation-contraction coupling but on rhythmic electrical activity generation.
Consequently, they do not display the long-lasting calcium influx-induced plateau. 
Their rhythmic activity is supported by a pacemaker funny current that conducts both $\mathrm{Na^{+}}$ and $\mathrm{K^{+}}$ and is coupled to intracellular $\mathrm{Ca^{2+}}$-clocks and membrane clocks~\cite{lakatta_coupled_2010}.
Accordingly, cardiac action potential duration and shape are finely adapted to cell function: in pacemaker cells to rhythmically generate electrical impulses, and in atrial and ventricular myocytes to conduct the activation to neighboring cells and trigger contraction.

In neurons, action potentials have an order-of-millisecond duration (Fig.~\ref{fig:fig2}A, right), making them brief compared to the multiple-milliseconds delay introduced by axonal and synaptic transmission. 
Together with the variability of synaptic delays, this enables high-frequency firing and leads to presynaptic activity asynchronously reaching postsynaptic targets, fostering distributed activation among connected neurons.
The spike brevity results from a rapid $\mathrm{K^{+}}$-channel activation and $\mathrm{Na^{+}}$-channel inactivation following the $\mathrm{Na^{+}}$‑driven upstroke. 
These mechanisms mediate a fast repolarization~\cite{bean_action_2007,kandel_principles_2013}.
The rapid repolarization is typically followed by a more extended afterhyperpolarization phase.
During this phase, the membrane potential becomes more negative than the resting value (typically at $\SI{-70}{mV}$) because the delayed deactivation of $\mathrm{K^{+}}$ channels allows it to approach the $\mathrm{K^{+}}$ equilibrium potential ($\SI{-90}{mV}$). 
The duration of this phase plays a key role in shaping neuronal excitability, as it regulates the inter-spike interval and thereby determines whether neurons fire isolated spikes, regular spike trains, or bursts.
Excitability is further shaped by $\mathrm{Ca^{2+}}$ fluxes as they trigger neurotransmitter release and are involved in learning mechanisms~\cite{hiess_subcellular_2022}.
Therefore, neuronal action potentials, because of their brevity relative to synaptic delays, support distributed and asynchronous activation among connected neurons, while their afterhyperpolarization dynamics determine each neuron’s characteristic firing pattern.

% In summary, from the organ scale to the single-cell scale, the heart and brain have evolved to efficiently support their respective functions.
% In cardiac layers, nearest-neighbor, bidirectional, fast gap junctions on a quasi-2D sheet enable the coherent propagation of electric wavefronts driven by the regular inputs from pacemaker cells. 
% The long duration of cardiac action potentials relative to intercellular transmission times further promotes local synchronization, ensuring reliable muscle contraction and efficient blood pumping.
% In contrast, the brain features directed, plastic synapses on a hierarchical, modular network, leading to distributed activation driven by irregular single-cell activity. 
% The brief duration of neuronal action potentials compared to synaptic delays suppresses global synchronization, further favoring efficient information processing.
% Despite the above-mentioned functional and physiological differences, electric signals in both organs are generated at the single-cell level from the same fundamental ion‑fluxes, which follow highly nonlinear dynamics. 
% This shared biophysical basis underlies the success of reductionist models that capture single‑cell voltage dynamics and link them to tissue‑level function through cell‑to‑cell coupling and network architecture.  
% While we adopt a reductionist approach here, recent advances in ion‑channel physiology and auxiliary regulatory mechanisms highlight additional layers of regulation and open new directions for future research.

% \section{Basic building blocks: Electrically excitable cells}
\section{Excitable cells as nonlinear units}
% (Irene)
\label{sec:excitable}

The generation of action potentials can be modeled at different levels of abstraction (Fig.~\ref{fig:fig2}B), reflecting a trade‑off between biological realism and analytical tractability.
On the one hand, detailed, ionic models explicitly describe the flow of each ion species across the cell membrane. 
On the other hand, reduced phenomenological models condense these electrical processes into a few variables, enabling analytical insights into general principles of excitability.

Ionic modeling was pioneered by the seminal work by Hodgkin and Huxley~\cite{hodgkin_quantitative_1952}, who developed a data-driven framework that fitted nonlinear ion-channel dynamics to experimental data by modeling channels as voltage-dependent conductance for independent ion currents (see Fig.~\ref{fig:hodgkin-huxley}).
Their approach remains the conceptual backbone of modern ionic, cardiac models such as the ten-Tusscher model of human ventricular myocytes~\cite{tentusscher_model_2004} or the O’Hara-Rudy model~\cite{ohara_simulation_2011}. 
These explicitly include the intracellular $\mathrm{Ca}^{2+}$ machinery and auxiliary currents that shape the cardiac waveform. 
The original Hodgkin-Huxley model~\cite{hodgkin_quantitative_1952} is still widely used in neuronal modeling, but reduced formulations such as the Conner-Stevens~\cite{connor_prediction_1971}, Morris–Lecar~\cite{morris_voltage_1981}, Izhikevich~\cite{izhikevich_simple_2003}, or adaptive‑exponential integrate‑and‑fire~\cite{brette_adaptive_2005} models have been introduced to reduce computational complexity while retaining biological grounding. 
These models capture the diversity of firing patterns with far fewer variables while retaining the key voltage‑dependent dynamics of several electrophysiological classes of cells.

Reduced phenomenological models still capture the fundamental features of action potential generation, despite not being directly mappable to specific biophysical quantities. 
A notable early example is van-der-Pol's relaxation oscillation~\cite{vanderpol_heartbeat_1928}, which inspired early physical models of excitable nerves~\cite{bonhoeffer_activation_1948}.
A later, influential low-dimensional model of excitability is the Fitzhugh-Nagumo (FHN) model, derived from the ionic Hodgkin-Huxley equations~\cite{fitzhugh_impulses_1961, nagumo_active_1962} and incorporating the van-der-Pol oscillator as a limiting case.
The FHN model couples a fast activator variable to a slower recovery variable, producing an action-potential-like response with a refractory period (Fig.~\ref{fig:fig2}C, Methods). 
Its dynamics can be represented either by plotting the time evolution of the variables (Fig.~\ref{fig:fig2}C2), clearly displaying the characteristic action potential waveform of the fast variable, or as a phase portrait, in which the trajectory of both variables evolves in a two‑dimensional plane (Fig.~\ref{fig:fig2}C3). 
In this latter case, one also includes the local gradients (gray arrows) together with the nullclines --- the loci where either variable’s rate of change is zero (black solid line for the fast variable, black dotted line for the slow).
From the phase portrait, one can see
a) that small (subthreshold) perturbations stay within the attractive gradient field of the stable fixed point (purple line),
b) that sufficiently strong (suprathreshold) perturbations get amplified by the gradient field, initiating an action potential (gray line). The action potential shape can be understood by following the gradient field: An initial upstroke (I) is followed by a plateau (II), rapid repolarization (III), and hyperpolarization (IV). 
From the phrase portrait, one can see that during phases II--III, it is impossible or harder to initiate another action potential. These phases constitute the refractory period.
The FHN model can be tuned to operate in excitable, oscillatory, or bistable regimes~\cite{datseris_nonlinear_2022}, making it a convenient scaffold for analytical and numerical studies of wave propagation, synchronization, and phase transitions in collections of excitable cells.

In summary, models of action potential generation range from detailed ionic descriptions to low-dimensional phenomenological accounts, reflecting a fundamental trade-off between biological realism and analytical tractability.
While ionic models accurately reproduce biophysical mechanisms, reduced models condense them into a few dynamical variables in a way that captures the general principles of electrical excitability --- threshold behavior, refractoriness, and nonlinear amplification. 
The FitzHugh–Nagumo model, in particular, offers an analytically tractable representation of excitability and illustrates how action potentials arise from the interaction of fast activation and slow recovery processes.
Importantly, action potential generation at the single-cell level is only one aspect of physiological function.
To extend these models to organ-level dynamics, cellular coupling, network architecture, and input patterns must be included.

\section{Function from collective interactions}
%(Simon)
\label{sec:collective}
To extend single-cell models of action potential generation to models of organ-level function, the cellular coupling and specifics of network architecture and input pattern in the heart and brain have to be taken into account.
These features determine the collective dynamics that emerges as the foundation for the different functions of the organs.
Here, we want to illustrate how function follows from interaction in a reductionist point of view by simplifying the intricate details of the physiological organization (Sec.~\ref{sec:physiology}). 
We employ the same FHN single-cell dynamics for modeling representative dynamics one finds in the heart and the brain, and focus on how the interplay of network architecture and input pattern shapes the qualitative differences in their collective dynamics.

% heart
From this reductionist perspective, healthy heart dynamics arises from a periodic stimulus to a quasi-two-dimensional sheet of strongly and bidirectionally coupled cells. 
To mimic this setting, we use a Fitzhugh-Nagumo model with a low internal noise level and local diffusive coupling on a 2D lattice (see Methods, Fig.~\ref{fig:fig4}A1). 
The sheet-like architecture creates an excitable medium in which a suprathreshold impulse spreads rapidly and uniformly in all directions (Fig.~\ref{fig:fig4}B1 left inset).
Combined with a periodic input that excites the system on one of the boundaries, this results in rhythmic, coherent traveling waves of activation (Fig.~\ref{fig:fig4}C1 left). 
The sum of the collective activation displays a typical shape reminiscent of healthy ECG recordings (Fig.~\ref{fig:fig4}D1 left). 
This illustrates how the traveling waves of activity underlying coordinated, reliable heart muscle contraction follow from the strongly-coupled, sheet-like architecture and the periodic stimulus. 

% brain
In a similar spirit, healthy brain dynamics may be seen as the result of time-dependent, distributed stimuli across sensory modalities to a complex, sparse, and directed network of weak connections. 
Here, we model this with a directed Erdős-Rényi graph~\cite{erdds_random_1959} of FitzHugh-Nagumo units (Fig.~\ref{fig:fig4}A2). 
To demonstrate how network architecture and input pattern shape emergent dynamics, we purposely choose the same single-cell model as in the cardiac model above, including the same average degree $\langle k\rangle =8$. 
We only change the network architecture and let a moderately noisy input take the role of the external stimulus. 
This fosters asynchronous activity with irregular single-unit activation (Fig.~\ref{fig:fig4}C2 left) with the sum of the collective activity resembling healthy EEG recordings (Fig.~\ref{fig:fig4}E2 left).
This demonstrates how the brain’s distributed activity capable of information processing arises from the weak coupling on a complex network architecture and varying stimuli. 

% living organ dynamics is more involved
The above models are intentionally simplified for illustrative purposes. In contrast, the physiological dynamics of the heart and brain are far more complex (see Sec.~\ref{sec:physiology}) and can be represented by more detailed models. 
The heart, for instance, is a highly organized three-dimensional structure rather than a collection of independent two-dimensional layers
Moreover, cardiomyocytes display characteristic dependencies of the action potential duration and the conduction velocity on the activation period. 
These are absent or weak in the FHN model and similar two-variable models, but have been matched specifically in other models~\cite{fenton_vortex_1998} (see Fig.~\ref{fig:spatiotemp}). 
They can cause dynamical instabilities known as alternans, which have also received significant attention as drivers of arrhythmias~\cite{echebarria_amplitude_2007}. 
The living brain poses a complex, adaptive ---not random and static--- network. Further, biological neurons can be inhibitory~\cite{vreeswijk_chaos_1996, brunel_dynamics_2000}, interact with glial cells~\cite{perea_tripartite_2009, pannasch_astroglial_2011, allen_cell_2017}, and deplete their neurotransmitter vesicles~\cite{barbieri_irregular_2007, yamamoto_modular_2023}. 
The emergent dynamics in the brain is thus not random but features weakly correlated spike trains that generate broadband fluctuations encoding high-dimensional, time-varying representations (so-called neural manifolds) of sensory inputs, internal states, and motor plans~\cite{gallego_neural_2017}. 
Yet, while these mechanisms have to be considered in efforts that aim to directly compare models to medical observation, the salient features and differences of heart and brain collective dynamics are captured by the reductionist model.

\afterpage{
\newsavebox{\footnotemarksinus}
\sbox{\footnotemarksinus}{\footnotemark}
\footnotetext{dataset \urlSinus}
\newsavebox{\footnotemarkvf}
\sbox{\footnotemarkvf}{\footnotemark}
\footnotetext{dataset \urlVF}
\begin{figure}[t]
    \centering
    \includegraphics{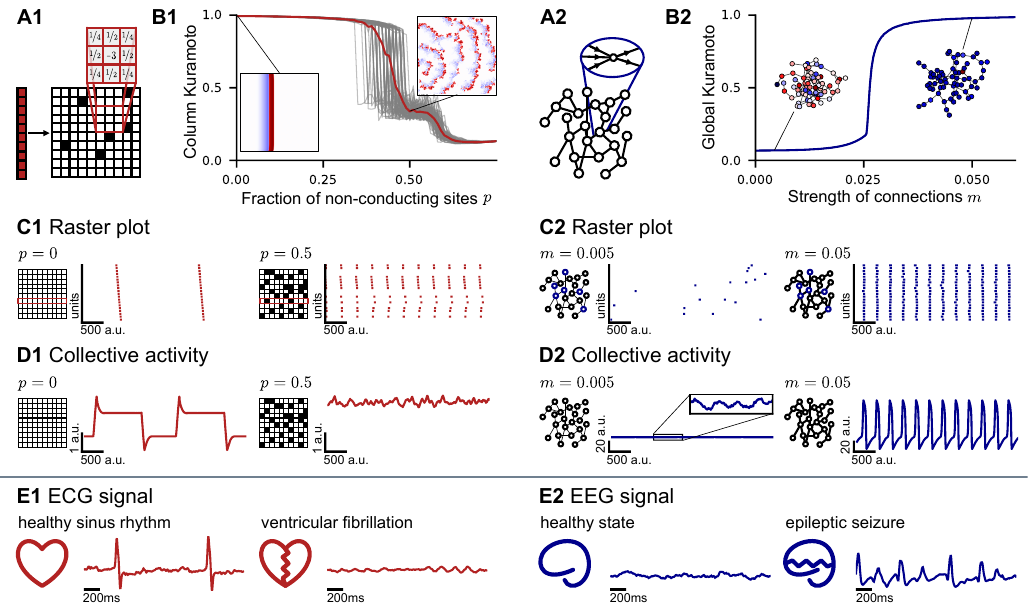}
    \caption{\textbf{Healthy heart and brain dynamics reside on opposite sides of a synchronization transition}. 
    \textbf{A1-D1}. Diffusively coupled FHN model on a 2D lattice to illustrate activity propagation in the myocardium.
    \textbf{A2-D2}. FHN model on a random network to illustrate activity propagation in the cortex.
    \textbf{B}. Synchronization order parameter, for the 2D lattice defined column-wise (see \methods). Insets show examples ($p=0$ and $p=0.5$ for the lattice and $m=0.005$ and $m=0.05$ for the random network).
    \textbf{C}. Raster plots for exemplary states (as in B) showing action potentials as bars for 20 selected units. 
    For the lattice, these units come from the central row only.
    \textbf{D}. Collective activity for exemplary states (as in B-C). 
    \textbf{E}. Examples of human ECG~\cite{manilo_ecg_2021, nemirko_ecg_2022, goldberger_physiobank_2000}, including a healthy sinus rhythm\usebox{\footnotemarksinus} and a ventricular fibrillation\usebox{\footnotemarkvf} (E1), and human EEG~\cite{andrzejak_indications_2001}, including a healthy state (dataset N062) and an epileptic seizure (dataset S056) state (E2).
    }
    \label{fig:fig4}
\end{figure}
}

Thus, the same elementary excitable units give rise to two opposite functional regimes solely because of differences in cell-to-cell coupling, network architecture, and input pattern.
A locally and bidirectionally coupled, sheet‑like architecture with a periodic input yields rhythmic, coherent traveling waves that underlie the heart’s mechanical pumping, whereas a directed complex network with noisy input produces irregular activity with low coherence that supports the brain’s information processing and input encoding. 
This dichotomy provides a natural framework for understanding how perturbations that shift the balance of synchrony may lead to pathological states.

%%%%%%%%%%%%%%%%%%%%%%%%%%%%%%%%%%%%%%%%%%%%%%%%%%%%%%%%%%%%%%%%%%%%%%%%%%%%%%%%%%%%%%%%%%%%%%%%%%%%%%%%%
\section{Function and Malfunction as two sides of a synchronization phase transition}
\label{sec:malfunction}

While healthy heart and brain function rely on the presence and lack of synchronization, respectively, pathologies can be interpreted as switches between these states. 
Asynchronous activity in the heart fails to support coordinated contractions, while excessive synchrony in the brain disrupts information processing.
From the viewpoint of nonlinear dynamics, the function and malfunction of both organs can be cast onto a single picture: a population of coupled excitable units may reside either in a synchronized state or in an asynchronous state, and varying a control parameter can drive a transition between them. 
Here, we illustrate how pathologies arise in the heart and the brain, and demonstrate, within the framework of elementary excitable units (FHN), how those macroscopic switches in synchrony can arise from changes in the effective coupling between cells.

%heart physiology
In the heart, interruptions of the synchronized action potential wavefront can lead to spatially uncoordinated or absent contractions, resulting in inefficient or absent blood pumping~\cite{prystowsky_cardiac_2020}. 
Such interruptions can arise from local defects, which can originate from the buildup of excessive connective tissue (fibrosis or scar tissue). 
These structural microscopic defects can disrupt the coherent action potential wavefront, leading to spiral wave formation (anatomical reentry)~\cite{dejong_fibrosis_2011, nattel_atrial_2008, shajahan_scanning_2016} or the termination of the electric signals (asystole).
The breakdown of the synchronized wavefront can also be caused by alterations in cell-intrinsic connexome organization or ion-channel properties, leading to pathological collective dynamics such as functional reentry. 
These alterations include lateralization of connexin hemichannels to subcellular regions outside the intercalated disc~\cite{nielsen_intercalated_2023, contreras_functioning_2003}. 
Such lateralization creates unpaired hemichannels on the cell membrane, which can open and flood the cell with calcium ions. 
The resulting intracellular calcium overload alters action potential generation and shape. 
A similar overload can also be caused by abnormal intracellular calcium dynamics due to impaired buffering~\cite{fakuade_impaired_2024, eisner_physiology_2023}.  
The pathological loss of synchronization is evident in physiological recordings: ECGs of malfunctioning hearts (Fig.~\ref{fig:fig4}E1) display fluctuating, aperiodic collective activity. 
Thus, while a healthy heart features synchronized traveling wavefronts along the intended conduction path, pathological, arrhythmic states arise when this coherence is lost, and the activity becomes asynchronous.

%heart modelling
To imitate pathological cardiac conditions in the 2D FHN lattice model, we randomly turn lattice units into non-conducting defects with probability $p$.
These structural microscopic defects are a simple model of fibrosis, scar tissue, or any localized impediment to activity spread that can give rise to cardiac arrhythmia. 
We quantify the coherence of wave propagation using a column-wise Kuramoto order parameter (see \methods and Ref.~\cite{kuramoto_selfentrainment_1975}), which measures phase alignment within each column of the lattice. 
In the healthy case, the wavefront travels from left to right and simultaneously activates all units at the same horizontal position, so that the column-wise order parameter captures propagation coherence along this intended direction.
In the real heart, the analogous reference would be the physiologically determined conduction path through the myocardium, respecting the specialized conduction system and tissue geometry (see Sec.~\ref{sec:physiology}).
As $p$ is increased, this order parameter decreases (Fig.~\ref{fig:fig4}B1). 
Beyond a modest defect density, the order parameter drops abruptly (gray lines show results for individual realizations of defect patterns).
The collective activity consequently changes from a coherent traveling wave with regular single-unit activity at $p=0$ to a noisy trace at $p=0.5$ (Fig.~\ref{fig:fig4}B1-D1), akin to healthy and pathological ECGs (Fig.~\ref{fig:fig4}E1).
In the noisy regime, the activity is neither coordinated across the lattice nor driven by the pacemaker stimulus but displays self-sustained spiral waves (Fig.\ref{fig:fig4}B1, inset).
When $p$ approaches the percolation threshold ($p_c\approx0.59$ in 2D for uncorrelated defects~\cite{ziff_spanning_1992, malarz_squarelattice_2005}), the conductive network fragments, propagation ceases, and the system becomes silent, producing only asynchronous background noise.
These qualitative behaviors are consistent with more realistic models of cardiac dynamics in the presence of structural microscopic defects~\cite{tentusscher_influence_2007, alonso_reentry_2013, gouveadebarros_simulation_2015, bittihn_emergent_2017, biasi_diffuse_2022, molla_insight_2024, dix_influence_2024} up to the fact that the FHN model is known not to capture (transient) chaos in two dimensions.
Such chaotic trajectories appear in more realistic models for low defect densities, preceding the phases of self-sustaining spiral waves and silent dynamics that occur in a very disordered medium~\cite{dix_influence_2024}. 
Thus, decreasing the effective coupling can induce (transient) chaos, self-sustaining spiral waves, and, ultimately, the termination of activity. 

%brain physiology
In the brain, excessively synchronous spiking~\cite{lehnertz_synchronization_2009} leads to poor information encoding and processing. 
While mild synchronization underlies communication between different brain areas~\cite{palmigiano_flexible_2017, bastos_communication_2015}, excessive synchrony from unrestrained excitatory activation is a hallmark of epileptic seizures~\cite{schwartzkroin_origins_1997, sumadewi_biomolecular_2023}, and dysfunctional \textit{in vitro} neuronal cultures~\cite{wagenaar_extremely_2006, levina_subsampling_2017, zierenberg_homeostatic_2018}.
Synchronous firing reduces the variability and independence of neural responses, leading to a decrease in entropy as neural activity becomes more predictable and thus carries less information. 
Consequently, when excessively synchronous firing occurs, the population encodes about the same information as a single neuron, impairing information processing. 
Such synchronous spiking is evident in EEG recordings of epileptic patients, where typical waveforms known as ictal discharges appear (Fig.~\ref{fig:fig4}E2). 
These simultaneous action-potential discharges can amplify synchrony across the network by reinforcing excitatory feedback and recruiting neighboring neurons into synchronized firing. 
This unrestrained excitatory activation can disrupt the function of inhibitory neurons, which, alongside regulatory mechanisms (see Sec.~\ref{sec:plasticity}), counteract excessive synchrony in healthy conditions~\cite{scharfman_neurobiology_2007, hagemann_assessing_2021}.
Thus, conversely to the heart, a transition from desynchronized to strongly synchronized activity leads to pathological states for the brain that are accompanied by a reduction in encoding capacity.

%brain modelling
In the random network FHN model, we reproduce excessive excitatory activation by scaling the coupling strength $m$.
This adjusts how strongly nodes influence each other.
As $m$ increases, the nodes drive each other’s activity more strongly, and the global Kuramoto order parameter increases continuously (Fig.~\ref{fig:fig4}B2).
The collective signal consequently changes from input-driven asynchronous dynamics with irregular single-unit activity ($m=0.005$) to almost global synchronization ($m=0.05$).
Additionally, we find that the variety of states, quantified by pattern entropy $H$ (see \methods), is high for asynchronous irregular dynamics ($H=0.94$, low redundancy, efficient coding) and low for synchronous dynamics ($H=0.23$, high redundancy, inefficient coding), see SM for a full scan.
This is consistent with more realistic models of neural dynamics that exhibit phase transitions~\cite{brunel_dynamics_2000, munoz_colloquium_2018}, and is also similar to observations in more complex networks~\cite{moretti_griffiths_2013, munoz_colloquium_2018, yamamoto_modular_2023}. 
Thus, increasing the coupling strength leads to a transition from asynchronous, high-entropy dynamics, which are well-suited for complex computations, to synchronous, low-entropy dynamics, as seen in epileptic seizures.

%summary
To summarize, we illustrated how the functional states of a healthy heart and a healthy brain can be considered to reside on opposite sides of a synchronization phase transition. 
Further, we demonstrated how crossing this transition via an altered effective coupling strength can lead to dysfunction: 
Microscopic defects in the heart (due to, e.g., fibrosis) can lead to arrhythmic, asynchronous wave patterns and hence uncoordinated contractions. 
Conversely, unrestrained excitatory coupling in the brain can lead to epileptic, synchronous firing with low information encoding capacity. 
This raises the question of how the heart and brain adapt to external and internal changes to maintain their functional (de)synchronized states and which mechanisms they use for this purpose.

\section{Modulation and plasticity}
\label{sec:plasticity}

% To preserve their physiological function, b
Both heart and brain use diverse adaptation mechanisms that adjust their dynamics in response to internal or external changes. 
These mechanisms can induce metabolic, mechanical, electrical, and structural changes in cardiomyocytes, neurons, and their connections.
Depending on their persistence, adaptation mechanisms can be categorized as modulation or plasticity, although these terms are loosely defined and sometimes overlap in both cardiac and neural research. 
Modulation typically denotes transient, reversible changes in the functional properties of a bulk population of cells, whereas plasticity refers to more persistent alterations in physical structures, such as formation, pruning, or long‑term reweighting of cell‑to‑cell connections.
In this section, we review the principal mechanisms of modulation and plasticity observed in the heart and the brain.

%modulation
Typical modulation mechanisms differ between the heart and the brain.
Cardiac modulation is strongly influenced by external feedback from other organs.
A prominent example is the control of cardiac output (amount of blood pumped in a given time) via the cardioregulatory center of the autonomic nervous system, which monitors heart function (blood pressure and composition) through cardiac and aortic receptors and can regulate the heartbeat frequency and volume through the release of neurotransmitters (noradrenaline and acetylcholine) by the sympathetic and parasympathetic nervous system at the SA node, the atrioventricular node and atrial and ventricular tissues~\cite{klabunde_cardiovascular_2011, ripplinger_neural_2022}. 
Typically, noradrenaline is released at neuro-cardiac synapse-like junctions (sympathetic activation), increasing heart rate, shortening ventricular action potential duration, thus supporting faster wave propagation when cardiac output must rise.
Cardiac contractile forces can further be augmented during acute sympathetic activation (e.g., excitement, exercise, or fight-or-flight) via the increase of calcium-triggered calcium release from the sarcoplasmic reticulum\cite{ripplinger_neural_2022}.
% brain
In contrast, neural activity is mostly modulated by intrinsic feedback loops and neuromodulators~\cite{krishnaswamy_demystifying_2019}.  
An important example is short-term regulation of neural activity via synaptic resource depletion~\cite{tsodyks_neural_1997, jahne_presynaptic_2021}, which transiently reduces synaptic efficacy during sustained activation. 
This negative feedback constrains energy consumption and counteracts excessive synchronous activity.
Excessive synchronous spiking caused by unrestrained excitatory activation is further counteracted by activity-induced pH changes: Periods of intense activity can cause local and transient acidification (lower pH)~\cite{sinning_minireview_2013} that dampens neural activity by reducing synaptic gain and spike probability~\cite{magnotta_detecting_2012, chesler_modulation_1992}. 
In summary, modulation in the heart is largely mediated by extrinsic feedback, whereas in the brain it primarily involves internal feedback. 
In both organs, it helps to prevent pathological activity by adapting electrical excitation properties to external and internal demands.

%Plasticity
Typical plasticity mechanisms in the heart and the brain are more similar than modulatory adaptation, but important differences remain.
While neurons can flexibly form and adapt synaptic connections over time, cardiomyocytes are bound to the local architecture of the myocardium and are largely constrained to pre-existing cell–cell coupling within the tissue.
However, they can adjust the coupling strength to neighboring cells through changes in the expression and localization of connexin hemichannels~\cite{rodriguez-sinovas_connexins_2021} and through remodeling of connexome organization between adjacent cells~\cite{nielsen_intercalated_2023}. 
These changes can help to preserve functional signal propagation in case of chronic changes in blood pressure or flow, but may also contribute to disease progression. 
For example, some pathological conditions, such as the loss of neighboring cardiomyocytes or pathogenic external stimuli such as chronically elevated catecholamines, can lead to compensatory or maladaptive hypertrophic growth of cardiomyocytes~\cite{hill_cardiac_2008}.
This cellular enlargement helps, to a certain extent, to maintain the amount of blood pumped by the heart, but, when excessive, can contribute to heart failure or outflow-tract obstruction~\cite{tran_cardiac_2025}. 
Pathological hypertrophy is often accompanied by disruption of the canonical end-to-end gap junction organization, which may contribute to conduction abnormalities, electrical disturbances, and sudden cardiac death. 
In contrast to cardiomyocytes, neurons support learning by actively forming and pruning connections, as well as by adjusting their synaptic strength~\cite{dayan_theoretical_2001}. 
These changes occur through structural plasticity (growth and rewiring of dendrites and axons) or synaptic plasticity (tuning of coupling strength).
Synaptic plasticity can be broadly divided into two categories:
homeostatic plasticity~\cite{turrigiano_homeostatic_2004} and spike-time-dependent plasticity (STDP)~\cite{caporale_spike_2008}.
Homeostatic plasticity maintains the firing rates of individual neurons within a functional range, preventing both the unrestrained excitation that might lead to excessive synchrony and epileptic seizures (see Sec.~\ref{sec:malfunction}) and silence. 
However, homeostatic compensation upon visual deprivation can lead to pathological effects and has been linked to hallucinations, such as in the Charles Bonnet Syndrome~\cite{series_hallucinations_2010, reichert_charles_2013, deistler_tactile_2019}.
By contrast, the major mechanism underlying STDP is the modification of coupling strengths based on the timing of action potentials between pairs of neurons.
The precise timing rules of STDP can vary between brain areas and synapse types; for most excitatory glutamatergic synapses, STDP induces long-term potentiation (synaptic strengthening) when the presynaptic spike occurs shortly before the postsynaptic spike, and long-term depression (synaptic weakening) when the postsynaptic spike precedes the presynaptic spike.
In summary, plasticity mechanisms in both organs are activated to achieve and preserve function, but can also contribute to disease progression.

% shared pathways % to be discussed: open issues and inputs by WHZ and MPZ
Remarkably, many adaptation mechanisms and their signaling pathways are shared between cardiomyocytes and neurons, often mediated by their common surroundings: The extracellular matrix. 
In both organs, the ECM must balance structural integrity with dynamic remodeling to accommodate single‑cell and connectivity changes~\cite{bonnans_remodelling_2014}. 
Dysregulated ECM remodeling can cause malfunction: Excessive ECM proliferation can lead to fibrosis in both heart and brain~\cite{dorrier_emerging_2022} (see Sec.~\ref{sec:malfunction}), and changes in protein expression in the brain ECM are associated with neurodegenerative diseases and psychiatric disorders~\cite{dityatev_brain_2021}.
ECM remodeling typically involves protease‑mediated cleavage followed by synthesis of new matrix components.  
One example of an ECM-cleavage pathway shared between heart and brain is the protease MMP-9, which participates in normal cardiac and neural remodeling~\cite {rodrigues_role_2024} and has neuromodulatory effects in the brain, adapting synaptic dynamics through long-term potentiation (LTP).
Consequently, dysregulation in MMP-9 can cause maladaptive ECM remodeling: For example, in the brain, elevated MMP‑9 activity is implicated in vasculature damage, lymphoma, and the Fragile-X syndrome~\cite{shi_targeting_2024}, while in the heart, it has been associated with matrix degradation and pathological processes that contribute to heart failure~\cite{rodrigues_role_2024}.
Ultimately, a deeper understanding of the ECM's dynamics is crucial for elucidating both normal adaptation and pathological failure in heart and brain, and for developing therapies targeting ECM dysregulation. 

% conclusion
To summarize, the heart and brain employ diverse adaptation mechanisms to maintain their functional states; the failure of these mechanisms leads to organ malfunction. 
Adaptation strategies may be intrinsic to the organ or involve extrinsic feedback, including interactions between the cardiac and nervous systems.
Importantly, some operate similarly in both organs, even at the molecular level. 
Therefore, both fields stand to benefit from cross‑disciplinary knowledge exchange, shared terminology, and common methodologies.

\section{Open questions}
\label{sec:openquestions}
The study of both the heart and brain continuously opens questions.
Framing ---and potentially solving--- these questions for one organ may also yield valuable insights into the other, given their shared foundational principles (Sec.~\ref{sec:excitable}).
This foundation covers their shared physiological cellular mechanisms as well as the emergence of collective dynamics that can be captured by the nonlinear-dynamics perspective we have provided here.
We have illustrated how cardiac pathologies can be viewed as a shift away from the healthy, synchronized beating pattern to a more chaotic and asynchronous state, or how epileptic seizures can be seen as a result of excessive synchronization within specific brain regions. 
Understanding the factors that influence these transitions and the potential tipping points could be crucial for developing preventative measures and therapeutic interventions.

We believe that by fostering a cross-talk between the fields of cardiac and brain research, and by applying a nonlinear dynamics lens, we will be able to unlock deeper insights into the fundamental principles governing the function of heart and brain. 
The resulting unified approach holds immense promise for developing novel therapies and ultimately for improving human health and well-being.
We thus conclude our perspective by providing a list of open questions that we believe warrant future attention.

\textbf{Do heart and brain need to be viewed as a coupled dynamical system?} % merge from MPZ's contribution and our example
Alongside individual organ research, it is crucial to better understand brain–heart interactions. These interactions dictate how both organs influence and regulate each other's functioning - in health and disease.
These influences involve the extrinsic autonomic nervous system and the often-overlooked intrinsic cardiac nervous system (the heart’s "little brain"), located in the cardiac ganglia~\cite{franzoso_putting_2016}. These nervous systems are in direct contact with cardiomyocytes at the neurocardiac junction~\cite{mongillo_neurocardiac_2020, dokshokova_nerve_2022}. This junction displays synapse‑like organization with directed-specific chemical signaling both from neurons to cardiomyocytes and vice versa (see Sec.\ref{sec:plasticity}). However, the molecular architecture, plasticity rules, and disease‑induced remodeling of these "neurocardiac synapses" remain largely unresolved~\cite{mongillo_neurocardiac_2020, dokshokova_nerve_2022}.
Conversely, the structure of neuron-neuron connections in the extrinsic and intrinsic nervous systems is better understood. Like in the brain, they are made up of chemical synapses that undergo neural plasticity~\cite{wagner_aging_2023, ieda_endothelin1_2004}. Understanding how these interactions affect diseases is crucial. For instance, in cardiac pathologies, neural plasticity can further exacerbate heart malfunction. For example, after myocardial infarction or chronic pressure/volume overload, inflammatory signaling and axonal sprouting drive heterogeneous sympathetic hyperinnervation and autonomic circuit remodeling, creating spatially complex electrophysiological substrates that favor malignant arrhythmias~\cite{zhou_mechanisms_2004, cao_relationship_2000}.
These multiscale structural and functional changes unfold within a strongly nonlinear excitable medium, where small perturbations of autonomic input can trigger complex or chaotic cardiac dynamics~\cite{chay_impulse_1984,chay_phase_1985,tran_bifurcation_2009}. This close coupling between neural and cardiac dynamics, and in particular the poorly understood neurocardiac synapse and neurocardiac plasticity, poses a critical but still largely unmapped interface where disease‑driven remodeling and nonlinear brain-heart interactions interfere. 

\textbf{How do pathological failures propagate between heart and brain?} %contributed by DK
Several clinical syndromes show that the coupling between heart and brain can propagate diseases between one organ and the other~\cite{haque_heartbrain_2026}. For example, in Takotsubo syndrome, acute emotional or neurological stress can cause transient cardiac dysfunction, most likely through sympathetic overactivation, catecholamine-mediated injury, microvascular dysfunction, and altered excitation–contraction coupling~\cite{wittstein_sympathetic_2016}. Additionally, stroke, subarachnoid hemorrhage, traumatic brain injury, and epilepsy can provoke arrhythmias, repolarization abnormalities, myocardial injury, or sudden cardiac death. These examples illustrate how disturbed brain activity and autonomic control can destabilize cardiac dynamics. In the opposite direction, chronic heart failure and atrial fibrillation are associated with cognitive decline and dementia, potentially due to reduced cerebral perfusion, embolic events, inflammation, and shared vascular risk factors~\cite{kjeldsen_causal_2025}. These clinical examples raise the question of whether heart–brain diseases can be understood as failures of a coupled dynamical system, in which perturbations in one organ push the other across a pathological transition~\cite{scheitz_bidirectional_2026}. Such a view may guide therapies that stabilize inter-organ coupling rather than treating cardiac and neurological dysfunction in isolation.

\textbf{Do the heart and brain share conserved endocytosis-mediated ECM remodeling pathways?} %contributed by SL
Recently, ECM remodeling mechanisms independent of protease activity, and thus unrelated to MMP9 function (see Sec.~\ref{sec:plasticity}), have been identified in both cancer biology~\cite{sung_regulation_2011} and neuroscience~\cite{dankovich_extracellular_2021}. For metastasis to occur, cancer cells typically need to reshape the ECM to support their movement, in a process that involves the endocytosis and recycling of ECM components such as fibronectin. This occurs through an endosomal/lysosomal pathway, without the need for protease-driven degradation~\cite{sung_regulation_2011}. A similar protease-free process has been observed in the brain, where changes in the ECM structure to support synaptic plasticity can occur via endocytosis and recycling of ECM molecules, with Tenascin-R (TNR) being the first identified example~\cite{dankovich_extracellular_2021}. It is unclear whether similar mechanisms also contribute to cardiac ECM dynamics. The involvement of TNR-like molecules in heart ECM dynamics, including the glycoprotein Tenascin-C (TNC), makes this likely. TNC is upregulated in the heart under pathological conditions, including dilated cardiomyopathy, myocarditis, and myocardial infarction~\cite{imanaka-yoshida_tenascinc_2001}. In these contexts, TNC modulates cell-ECM interactions through processes known to depend on molecular endocytosis and recycling of surface molecules~\cite{paul_endocytic_2015}. This suggests that TNC dynamics, as for TNR in the brain, may be dependent on endocytosis and recycling.
To further investigate these dynamics, advanced imaging approaches that can follow molecules over long time periods will be required. This can be achieved with super-resolution shadow imaging (SUSHI)~\cite{tonnesen_superresolution_2018}, which introduces fluorophores in the intercellular space. Similar approaches, injecting or expressing ECM-binding molecules in living tissue, have been conceived and will be important in future heart and brain investigations~\cite{deangelis_tmem2_2017}.

\textbf{Can therapeutic strategies targeting ion channel dysfunctions in the brain inform therapies targeting analogous cardiac channelopathies, and vice versa?}
Calcium dynamics play a crucial role in both the heart and the brain, regulating cardiomyocyte contraction and neurotransmitter release, respectively (see Sec.~\ref{sec:physiology}). When these dynamics become dysregulated, they can give rise to pathological organ-level activity~\cite{santulli_mitochondrial_2015, pchitskaya_calcium_2018}. A striking example is the rare autosomal-dominant Timothy syndrome~\cite{jiang_current_2023}, where mutations in the gene \textit{CACNA1C} cause a conformational change in a type of calcium channel, causing prolonged calcium influx into cells. 
This leads, in the heart, to the prolongation of ventricular depolarization and repolarization, causing arrhythmias, bradycardia, atrioventricular block, and, in some cases, structural heart defects such as hypertrophic cardiomyopathy~\cite{yazawa_using_2011,dick_arrhythmogenesis_2016}. In the brain, it contributes to epilepsy, intellectual disability, and autism spectrum disorder. Remarkably, an antisense oligonucleotide therapy that showed promise in preclinical models targeting neural symptoms is now being tested for cardiac efficacy~\cite{chen_antisense_2024}. Such findings suggest that treatments developed for one organ may translate to the other for diseases rooted in shared ion channel mutations and dysfunctions.

\textbf{Is activity propagation in modern-day relatives of ancestral nervous systems similar to cardiac spatiotemporal patterns?} %contribution from AN
Our comparison between the excitable tissues of the mammalian heart and brain revealed a fundamental dichotomy. 
Cardiac cells are coupled to their immediate neighbors with a bidirectional, instantaneous electrical coupling. 
In contrast, cortical neurons mostly form long-range, directed, and delayed chemical connections.
Although this binary picture aids clarity, other animals display different neural coupling mechanisms that partially resemble cardiac coupling. 
For instance, Comb jellies (Ctenophores) are covered with nerve nets that rely on electrical connections for information processing~\cite{shelton_electrical_1982} and on chemical synapses only to innervate actuators~\cite{sachkova_neuropeptide_2021}. 
Ctenophores are regarded as modern-day relatives of the earliest animals with nervous systems. 
Connections between their nerve cells are bidirectional and electrical: their membranes are fully fused into a continuous network spanning the entire Ctenophore body~\cite{burkhardt_syncytial_2023}. 
Although the functional properties of ctenophore nerve cells are not yet fully understood, signal propagation within the nerve network can be expected to occur in spatiotemporal patterns more closely resembling those found in cardiac tissue rather than in cortical networks.

\textbf{What can we learn from other excitable cell tissues?} %contributed by AN
Excitable cells are found not only in the heart and brain but also in other tissues. 
For example, luteinizing hormone-releasing cells (LH cells) in the fish pituitary~\cite{golan_anatomical_2016}, when stimulated by another hormone, fire action potentials that trigger LH release. 
These cells are connected by gap junctions, forming a functional syncytium that excludes other cell types interspersed in the tissue.
Their heart-like coupling results in synchronized collective dynamics with coordinated oscillations that produce pulsatile LH release.
Many more examples of such networks exist, from astrocytes in the retina and cortex~\cite{newman_propagation_2001} to insulin-secreting islets of Langerhans in the pancreas~\cite{benninger_gap_2008}.
While some rely exclusively on intracellular coupling, others depend on the release of signaling molecules into the extracellular space, creating additional feedback pathways, as seen in several types of mammalian pituitary cells~\cite{le_tissier_updated_2017}. 
%These examples show that nature employs excitable cells not only as ``synchronized hearts'' or ``chaotic brains,'' but also with intermediate dynamics or even switching between these states.
%Understanding the relationship between microscopic structure, collective dynamics, and function could help us reveal how excitable tissues generate robust and flexible physiological functions.

In this broader view, the heart and brain are not isolated examples, but two instructive limits of a wider design space, in which evolution tuned excitable cells, their interactions, and their adaptive mechanisms to generate the collective dynamics required for biological function.
Taken together, we believe that such a holistic perspective is not only conceptually illuminating but may also be practically useful: mechanisms that stabilize synchrony in one system can inspire questions about how synchrony is avoided, controlled, or restored in the other. 
More generally, insights gained in one excitable system may help identify organizing principles, failure modes, and intervention strategies in another.

%%%%%%%%%%%%%%%%%%%%%%%%%%%%%%%%%%%%%%%%%%%%%%%%%%%%%%%%%%%%%%%%%%%%%%%%%%%

\section*{Acknowledgements}
We want to thank Andreas Schneider and Maximilian Schäfer for the valuable discussion and comments on the text.
This research was conducted within the Max Planck School Matter to Life, supported by the Dieter Schwarz Foundation in collaboration with the Max Planck Society.
This work was funded by the Deutsche Forschungsgemeinschaft (DFG) via the MBExC under Germany’s Excellence Strategy-EXC 2067/1-390729940 and the SFB 1690/1.

\section*{Data Availability}
Experimental data reviewed in the figures are cited where displayed. Additional data generated from simulations can be reproduced with our GitHub repository\cite{PriesemannGroupHeartBrainMisc}

\par
\vspace{0.5\baselineskip}
\begin{center}
    \rule{0.5\textwidth}{0.5pt}
\end{center}
\par
\vspace{0.5\baselineskip}
\section*{Methods}
%\label{sec:methods}
\begingroup 
\smallskip

\textbf{FitzHugh-Nagumo model (FHN):} The FHN is a 2-variable phenomenological model of excitability with a fast variable $V$, which can be interpreted as the membrane potential, and a slow recovery variable $W$. 
For a network of excitable cells coupled by the matrix $\{J_{ij}\}$, we model the dynamics of cell $i$ as 

\begin{align}\label{eq:fhn}
    \frac{\mathrm{d}V_i}{\mathrm{d}t} &= a V_i (V_i-b)(1-V_i) - W_i + \sum_j J_{ij} V_j + I_i(t)+ \eta_i,\\
    \frac{\mathrm{d}W_i}{\mathrm{d}t} &= \epsilon (V_i-W_i),
\end{align}
where each cell receives a (time-dependent) input $I_i$ and additive, uncorrelated Gaussian noise $\eta_i$ to capture, e.g., fluctuations from variable synaptic transmission. $a$ relates to the plateau width in the action potential, $b$ tunes the excitation threshold, and (a small) $\epsilon$ realizes the time separation between the two variables and is proportional to the action potential duration.
We integrate these equations numerically using the Euler-Maruyama scheme. The code can be found in the educational notebook~\cite{pellini_educational_2026}.
For the following reductionist models, we use the same parameters for the individual excitable cell: $a=3$, $b=0.05$, and $\epsilon=0.01$.

%heart model
To mimic cardiac activity propagation, we simulate a $N=200\times 200$ 2D lattice with no charge flow allowed across the boundaries. 
The coupling matrix $J_{ij}=-D_V L_{ij}$ is defined by the Laplacian matrix $L_{ij}$ multiplied by the diffusion coefficient $D_V=0.04$. The Laplacian
$L_{ij}$ is built using the convolution of a nine-point stencil: \[
L =
\begin{pmatrix}
\frac14 & \frac12 & \frac14\\
\frac12 & -3 & \frac12\\
\frac14 & \frac12 & \frac14
\end{pmatrix}.
\]
We use no flux boundary conditions.
To minimize noise-induced action potentials, the noise's standard deviation is set to $\sigma^2_{\eta}=0.0001$. The periodic heartbeat is simulated by applying an external input $I_i=0.1$ to periodically trigger an action potential in the cell at the left boundary of the simulation frame. 
To prevent the simultaneous occurrence of two fronts in the lattice and to ensure long inter-beat times, the stimulus period is set to $2000=20/ \epsilon$, so about 7 times the action potential duration.
To achieve the transition from coherent traveling waves to broken waves and a noisy steady state, we simulate the presence of microfibrosis, following Refs.~\cite{alonso_reentry_2013, alonso_reentry_2016, gouveadebarros_simulation_2015} we introduce a probability $p$ with which each site is turned into a non-conductive unit. A non-conductive unit is uncoupled from the network dynamics and treated in the same way as the boundaries of the frame. 

%brain model
To illustrate brain-like activity propagation, we simulate $N=40000$ units coupled via a directed Erdős-Rényi graph~\cite{erdds_random_1959} with a fixed mean out-degree $k=8$. 
We set the coupling matrix to $J_{ij}= m/k A_{ij}$, where $A_{ij}$ is the adjacency matrix of the network and $m\geq 0$ sets the strength of the excitatory coupling. 
The noise variance is set to $\sigma^2_{\eta}=0.008$ so that it can trigger noise-induced action potentials. 
The external input is set to zero for all nodes.
To obtain a transition from random dynamics to strong synchronization, we directly tune the coupling strength $m$. 

\textbf{Synchronization measure:} The degree of synchronization over time is quantified using the real part of the Kuramoto order parameter~\cite{kuramoto_selfentrainment_1975}:
\begin{align}
r(t)e^{i \psi(t)}=\frac{1}{N}\sum_{k=1}^N e^{i \theta_k(t)}.
\end{align}
The phases $\theta_k$ are calculated from the Hilbert transform $\mathcal{H}$ of $V_k$ \cite[pp. 362--370]{pikovsky_synchronization_2001}.  The instantaneous phase of the signal is extracted as  
$\theta_k(t)=\arg(i\mathcal{H}[V_k](t)+V_k(t))
$. By averaging $r(t)$ over time, we obtain the global synchronization level: 
$R= \left\langle r(t)\right\rangle_t$

In the case of the 2D lattice, for $p=0$, we expect sites belonging to the same column to be perfectly synchronized as they are simultaneously reached by the periodic excitation, which travels horizontally.
Therefore, we measure synchronization column-wise, quantifying wave propagation coherence relative to this intended left-to-right direction: For each column $j$, we calculate $R_j$ disregarding the contribution of the defects and extract the total order parameter as $R=1/\sqrt{N}\sum_{j=1}^{\sqrt{N}} R_j(t)$.  

\textbf{Entropy measure:} 
We use a binary representation of the system's time evolution to quantify entropy. 
First, we define binary activity variables $A_i(t)$ from the time evolution of each $V_i$, considering a node active when its membrane potential exceeds a threshold $V_\mathrm{thres}$. We set $V_\mathrm{thres}=0.08>b$ for both organs so that an active node is in the depolarization phase of the action potential.
Next, we consider smaller subsets of $A_i(t)$ at fixed $t$. 
For each subset, we count over time the number of occurrences $n_i$ of each of the $\mathcal{N}$ possible binary patterns it can store and define the probability of the $i^{th }$ pattern as $p_i=(n_i+1)/(T+\mathcal{N})$, where $T$ is the total number of time-steps. 
This definition assumes a uniform probability distribution for patterns that are possible but not observed, thus counteracting the bias introduced by observing only a small number of patterns. 
Finally, we quantify the information content in the system by computing the Shannon entropy as $H=-\sum_{i}p_i \log_2 (p_i)$. 
\endgroup

%\section{Appendix}
\begin{bbb}[label=box:hodgkin]{Hodgkin-Huxley Model of Action Potential Generation}

The first measurement of the action potential in a nerve cell and its quantitative explanation through ionic currents was a major breakthrough for neuroscience, leading Alan Lloyd Hodgkin and Andrew Fielding Huxley to the 1963 Nobel Prize in Physiology or Medicine~\cite{_nobel_1963} (shared with Sir John Carew Eccles for complementary discoveries on synaptic transmission). 
Starting in 1938~\cite{hodgkin_action_1939}, they conducted a series of experiments on the electrical signal generation in a squid giant axon, which ultimately culminated in the landmark publication of their theory in 1952~\cite{hodgkin_quantitative_1952}.
 
Their theory concludes that the nerve cells' potential is controlled by the concentration difference of sodium and potassium ions (with other ionic species contributing mainly through leak currents) inside and outside of the cells' membrane. The membrane thus acts as an electrical capacitor, separating the charges and allowing the flow of ionic currents through gated ion channels.
The theory can be cast in the form of an equivalent electrical circuit (Fig.~\ref{fig:fig2}B).
Any external input current $I(t)$ enters the cell through the membrane capacitance $C_m$, potassium channels with conductance $g_K$, sodium channels with conductance $g_{Na}$, or through leaky channels $g_L$, which bundle other ions of lower importance in the squid axon. 
In each of the channels, a battery in the equivalent circuit mirrors the ions' reversal potential $E_{Na/K/L}$, i.e., the Nernst potential generated by the equilibrium concentration difference of the ions across the cell membrane~\cite{gerstner_neuronal_2014}. Kirchhoff's current law applied to the membrane then leads to a differential equation for the membrane potential $V(t)$~\cite{hodgkin_quantitative_1952},
\begin{equation}\label{eq:hodgkin-huxley-a}
    C_m \frac{\mathrm{d}V}{\mathrm{d}t}=-g_{Na}(V-E_{Na})-g_{K}(V-E_{K})-g_{L}(V-E_{L}) + I(t).
\end{equation}

Additionally, Hodgkin and Huxley experimentally found that the sodium and potassium conductances varied with membrane potential $V$. 
These changes in conductances arise from voltage-dependent gating processes that modulate the probability of ion channel opening. Accordingly, the activation of potassium channels can be modeled with the introduction of another voltage-dependent variable $n$ according to $g_K=\bar g_K n^4$. 
Similarly, sodium channels are controlled by two voltage-dependent variables $m$ and $h$ via $g_{Na}=\bar g_{Na} m^3h$, where $m$ activates the sodium channel, and $h$ inactivates it. 
Leak conductance, instead, is approximately constant $g_L=\bar g_L$. $n$, $m$ and $h$ all depend on the membrane potential with dynamics following~\cite{hodgkin_quantitative_1952}
\begin{align}
    \frac{\mathrm{d}n}{\mathrm{d}t}&=-\frac{1}{\tau_n(V)}(n-n_0(V))\label{eq:hodgkin-huxley-b}\\
    \frac{\mathrm{d}m}{\mathrm{d}t}&=-\frac{1}{\tau_m(V)}(m-m_0(V))\label{eq:hodgkin-huxley-c}\\
    \frac{\mathrm{d}h}{\mathrm{d}t}&=-\frac{1}{\tau_h(V)}(h-h_0(V)).\label{eq:hodgkin-huxley-d}
\end{align}
    \centering
    \includegraphics[width=\linewidth]{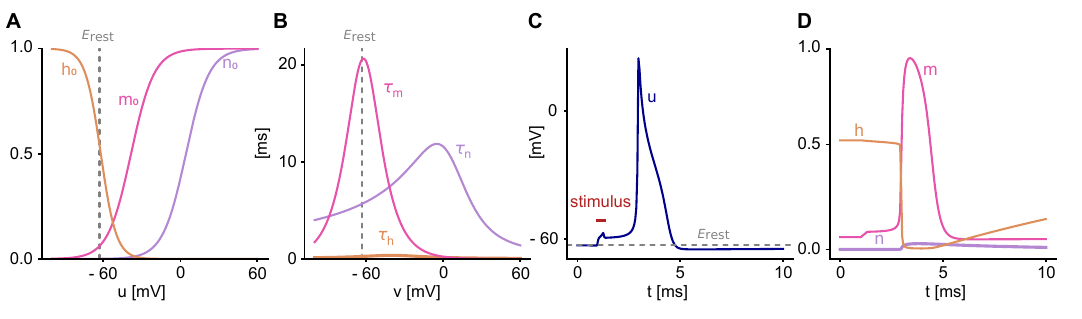}
    \captionof{figure}{\textbf{Action potential generation in the Hodgkin-Huxley model.} The dynamics of the voltage gates of sodium and potassium channels are mediated by Eq.~\eqref{eq:hodgkin-huxley-b}--\eqref{eq:hodgkin-huxley-d} together with the voltage-dependent equilibrium values (\textbf{A}) and timescales (\textbf{B}), adapted from~\cite{gerstner_neuronal_2014, mainen_model_1995, op_improved_1981}. (\textbf{C}), (\textbf{D}) Numerical simulation of an action potential generation following an initial stimulus. Here, $E_{Na}=55mV$, $E_K=-77mV$, $E_L=-65mV$, and $E_{rest}=-63mV$, $\bar g_{Na}=\SI{40}{mS/cm^2}$, $\bar g_K=\SI{35}{mS/cm^2}$, and $\bar g_L=\SI{0.3}{mS/cm^2}$, and $C_m = \SI{1}{\mu F/cm^2}$, all as reported in~\cite{gerstner_neuronal_2014}.}
    \label{fig:hodgkin-huxley}
\justifying

The voltage-dependendent equilibrium values ($n_0(V)$, $m_0(V)$, $h_0(V)$) and the time scales on which they are approached ($\tau_n(V)$, $\tau_m(V)$, $\tau_h(V)$) are given by the empirically determined functions in Fig.~\ref{fig:hodgkin-huxley}A, B. Together, the dynamics of $V$, $n$, $m$, and $h$ can describe the formation of an action potential: At the resting state of the membrane potential, both the potassium and the sodium channels are closed, i.e., $n$ and $m$ are close to zero while $h$ is on the order of (but usually well below) one.
If an external current triggers an increase in the potential over a certain threshold, $m$ starts to increase. 
This activates the sodium channels, letting $\mathrm{Na^+}$ flow into the cell, causing a positive feedback which further depolarizes the membrane and initiates the action potential. 
With a little delay, the potassium channels also activate via a rise in $n$ (Fig.~\ref{fig:hodgkin-huxley}A, B), letting $\mathrm{K^+}$ flow out of the cell, which curbs the sharp initial increase in potential. 
Subsequently, $h$ drops, inactivating the sodium channels such that the active potassium currents lower the membrane potential again, eventually hyperpolarizing the membrane below the resting state. 
Until the sodium channels recover from inactivation and potassium conductance decays, the membrane enters an absolute and then relative refractory period, during which sodium channel inactivation and elevated potassium conductance suppress or impede action potential initiation, respectively. 
\end{bbb}

\begin{bbb}[label=box:spatiotemporal]{Spatiotemporal dynamics of the heart}
\begin{figure}[H]
  \centering
  \includegraphics[width=\textwidth]{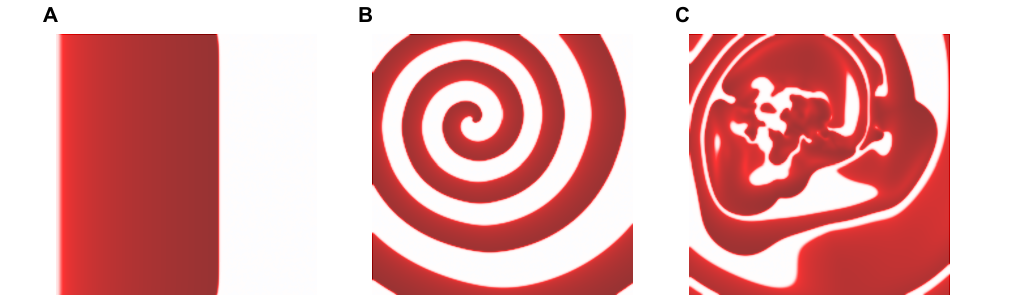}
  \caption{\textbf{From regular to irregular spatiotemporal dynamics of the heart.} The figures are obtained using the Aliev-Panfilov model~\cite{aliev_simple_1996}: $ \frac{dV}{dt} = a V (V-b)(1-V) - WV + D_V \nabla^2 V ,  \frac{dW}{dt} = \epsilon(V,W)(-W-cV(V-c-1))$ with $\epsilon(V,W)=\frac{\epsilon_0+\mu_1W}{\mu_2+V}$. As for the FitzHugh-Nagumo model, $V$ and $W$ are dimensionless variables.
\textbf{A.} Traveling wave corresponding to a functional, healthy state.
\textbf{B.} Spiral wave formation corresponding to reentry and arrhythmia. 
\textbf{C.} Spatio-temporal chaos corresponding to the onset of fibrillation, in this example due to increased excitability (i.e., decrease in the excitation threshold $b$) leading to spiral wave breakup.
Parameters (\textbf{A}--\textbf{C}): $a=8,\;\mu_1=0.01 ,\; \mu_2=0.3, \;c=0.1,\; \epsilon_0=0.01,\;D_V=0.3$. In panels (\textbf{A}), (\textbf{B}),  $b=0.2$, whereas $b=0.05$ in (\textbf{C}). In all panels, color indicates the excitation variable $V$.}
\label{fig:spatiotemp}
\end{figure}
Excitable media, such as cardiac tissue, exhibit diverse spatiotemporal activation patterns including periodic traveling waves, spiral waves (rotors), and spatiotemporal chaos (fibrillation), as illustrated in Fig.~\ref{fig:spatiotemp}A-C using the Aliev-Panfilov model~\cite{aliev_simple_1996,datseris_nonlinear_2022}. This model is a modification of the diffusively coupled FitzHugh-Nagumo model used in the main text that more faithfully captures some of the features of cardiac excitations, such as the shape of the action potential and so-called restitution properties (see Sec.~\ref{sec:collective}). The latter determines how the action potential duration and wave conduction velocity depend on the time passed since the last excitation and has been shown to be important to correctly describe the onset of arrhythmia in homogeneous tissue~\cite{karma_electrical_1994,fenton_vortex_1998,kim_action_2002}. In such systems, all of the dynamics mentioned above can be observed without local parameter variations or structural defects like those considered in the main text.

Periodic traveling waves (Fig.~\ref{fig:spatiotemp}A) represent an idealized healthy state of cardiac excitation propagation. These waves coherently propagate through the tissue, ensuring coordinated contractions essential for efficient heart function.

Spiral waves or rotors in homogeneous media (Fig.~\ref{fig:spatiotemp}B) correspond to functional reentry (see Sec.~\ref{sec:malfunction}), since there is no anatomical structure around which the activity circles. Functional reentry can develop out of a broken traveling wave: Rapid recovery allows previously excited regions to become excitable again quickly, enabling the wavefront to circle back onto itself and form stable spiral waves. Similarly, sufficiently slow conduction increases the time available for recovery ahead of the wavefront and can likewise promote reentry. Broken wave fronts, in contrast, are often exacerbated by heterogeneous recovery, as wavefronts collide with refractory regions in some locations but not others. Numerous modeling studies and reviews have examined the conditions controlling spiral wave dynamics, highlighting how changes in ionic currents, tissue refractoriness, and conduction velocity modulate rotor formation and stability (e.g.,~\cite{guevara_phase_1981, fenton_multiple_2002, pandit_rotors_2013, krogh-madsen_nonlinear_2012}).

The transition to spatiotemporal chaos (Fig.~\ref{fig:spatiotemp}C) associated with fibrillation occurs when the balance between excitation and recovery is significantly disrupted. A lowered excitation threshold ($b$), for instance, makes the tissue more susceptible to frequent, irregular activation from small perturbations. This destabilizes spiral waves, causing fragmentation into multiple wavelets that interact chaotically. Mechanisms underlying spiral wave breakup include spiral tip instability, meandering through Hopf bifurcations, and complex period-doubling cascades~\cite{barkley_linear_1992, karma_meandering_1990, fenton_multiple_2002, cherry_visualization_2008, krogh-madsen_nonlinear_2012, bittihn_complex_2014}. Wave propagation and arrhythmogenesis are also significantly influenced by structural features such as fiber orientation and its systematic rotation through cardiac tissue, as well as the specialized conduction system outside the bulk muscle~\cite{fenton_vortex_1998, bordas_rabbitspecific_2011, hooks_laminar_2007}. Other structural alterations, like localized parameter differences, fibrosis, or localized conduction obstacles, further alter the dynamics of reentry and fibrillation~\cite{tentusscher_influence_2007, alonso_reentry_2013, gouveadebarros_simulation_2015, shajahan_scanning_2016, bittihn_emergent_2017, biasi_diffuse_2022, molla_insight_2024, dix_influence_2024}.

We thus see that model analysis using the tools of nonlinear dynamics allows us to recover the key spatiotemporal patterns of cardiac pathologies and understand the causes underlying their formation. Similar tools can be used to devise effective treatments for these pathologies and test their effects in silico~\cite{hussaini_efficient_2024, luther_lowenergy_2011}. Thus, theoretical and computational efforts are a crucial component in understanding cardiac dynamics and developing effective treatments for pathological heart conditions. 

\end{bbb}

\bibliographystyle{unsrt}
\bibliography{heart-and-brain}

\end{document}